\documentclass[revtex4]{emulateapj}

\usepackage{amssymb}
\usepackage{amsmath}
\usepackage[]{graphicx}
\usepackage{enumerate}
\usepackage{subeqnarray}
\usepackage{cases}
\usepackage{mathrsfs,amssymb}

\newcommand{\hh}{\mathrm{H_{2}}}

\tightenlines

\begin{document}

\title{Interstellar dust charging in dense molecular clouds: Cosmic ray effects}

\author{A. V. Ivlev$^1$, M. Padovani$^{2,3}$, D. Galli$^3$, and P. Caselli$^1$}
\email[e-mail:~]{ivlev@mpe.mpg.de} \affiliation{$^1$Max-Planck-Institut f\"ur Extraterrestrische Physik, 85741 Garching,
Germany\\ $^2$ Laboratoire Univers et Particules de Montpellier, UMR 5299 du CNRS, Universit{\'e} de Montpellier, 34095
Montpellier, France\\ $^3$INAF-Osservatorio Astrofisico di Arcetri, 50125 Firenze, Italy}

\begin{abstract}

The local cosmic-ray (CR) spectra are calculated for typical characteristic regions of a cold dense molecular cloud, to
investigate two so far neglected mechanisms of dust charging: collection of suprathermal CR electrons and protons by grains,
and photoelectric emission from grains due to the UV radiation generated by CRs. The two mechanisms add to the conventional
charging by ambient plasma, produced in the cloud by CRs. We show that the CR-induced photoemission can dramatically modify
the charge distribution function for submicron grains. We demonstrate the importance of the obtained results for dust
coagulation: While the charging by ambient plasma alone leads to a strong Coulomb repulsion between grains and inhibits
their further coagulation, the combination with the photoemission provides optimum conditions for the growth of large dust
aggregates in a certain region of the cloud, corresponding to the densities $n(\hh)$ between $\sim10^4$~cm$^{-3}$ and
$\sim10^6$~cm$^{-3}$. The charging effect of CR is of generic nature, and therefore is expected to operate not only in dense
molecular clouds but also in the upper layers and the outer parts of protoplanetary discs.
\end{abstract}


\maketitle

\section{Introduction}

Interstellar dust grains in dense molecular clouds are subject to several electric charging processes
\citep[e.g.,][]{Draine1979,Draine1987,Weingartner2001}. The resulting net electric charge carried by micron or sub-micron
size grains has important consequences for the chemical and dynamical evolution of molecular clouds: it affects the process
of dust coagulation \citep[][]{Okuzumi2009,Dominik2007}, the rate of grain-catalyzed electron-ion recombination
\citep[][]{Mestel1956,Watson1974}, the amount of gas-phase elemental depletion \citep[][]{Spitzer1941}, and the electrical
resistivity of the cloud's plasma \citep[][]{Elmegreen1979,Wardle1999}. The resistivity, in turn, controls the coupling
between the neutral gas and the interstellar magnetic field, and eventually the dynamics of gravitational collapse of
molecular clouds and the formation of stars \citep[e.g.,][]{Nakano2002,Shu2006}.

Collisions of dust grains with the plasma of thermal electrons and ions from the gas (hereafter, cold plasma charging)
represent an important dust charging process in molecular clouds \citep[e.g.,][]{Draine1987,Draine2011Book}. Since electrons
of mass $m_e$ have a thermal speed which is much larger than that of ions of mass $m_i$ (by the factor
$\sqrt{m_i/m_e}\gg1$), grains acquire by this process a (predominantly) negative charge. The photoelectric effect (also
called photoemission), on the other hand, results in positive charging of dust grains, and is set by the radiation field in
the cloud at energies above a few eV. Photoemission is an important charging process for diffuse gas with visual extinction
$A_{\rm V}\lesssim10$ \citep[e.g.,][]{Bakes1994,Weingartner2001}. As the interstellar radiation field is exponentially
attenuated with increasing $A_{\rm V}$, photoemission is usually neglected to compute the charge distribution of grains in
the dense gas of molecular cloud cores \citep[e.g.,][]{Umebayashi1980,Nishi1991}.

Cold plasma charging and photoemission are usually assumed to be the dominant grain charging mechanisms in the cold
interstellar medium. In this paper we study the effects of cosmic rays (CRs) on the charging of submicron dust grains in
molecular clouds. We focus on two charging processes that contribute in addition to the cold-plasma charging, but have been
neglected so far. By calculating the local CR spectra for typical cloud regions, we investigate the effects of ({\em i}\/)
collection of suprathermal CR electrons and protons by grains \citep[][]{Shchekinov2007} and ({\em ii}\/) photoelectric
emission from grains due to the UV radiation generated by CRs. Using the cold-plasma collection as the ``reference case'',
we show that the photoelectric emission can dramatically modify the charge distribution function for dust in almost the
entire cloud, and discuss important implications of the obtained results. In particular, we point out that while the
cold-plasma charging alone leads to a strong Coulomb repulsion between grains and inhibits their further coagulation, the
combination with the CR-induced photoemission provides optimum conditions for the growth of large dust aggregates in a
certain region of the cloud.

\section{CR properties relevant to dust charging}\label{mechanisms}

The specific intensities (or spectra) of CR protons and electrons inside a dense molecular cloud are determined by the
interstellar CR spectra. In order to constrain the trend of the interstellar spectra at high energies ($E\gtrsim500$~MeV),
we use the latest results of the Alpha Magnetic Spectrometer (AMS-02), mounted on the International Space Station
\citep[][]{Aguilar2014,Aguilar2015}. The high-energy spectrum slope is $-3.2$ for electrons, while for protons it is $-2.7$.

At lower energies, the shape of the interstellar CR spectrum is highly uncertain due to the effect of Solar modulation
\citep[see, e.g.,][]{Putze2011}. Data collected by the Voyager~1 spacecraft from a region beyond the Solar termination shock
are extremely useful in this context \citep[][]{Webber1998}, as they provide a lower limit to the CR spectrum in the energy
range from $\sim5$~MeV to $\sim50$~MeV that should be not too different from the actual interstellar value
\citep[][]{Stone2013}. However, two caveats should be kept in mind. First, Voyager~1 has not yet entered interstellar space,
as the magnetic field detected by magnetometers on board the spacecraft still retains some characteristics of the Solar wind
magnetic field \citep[][]{Burlaga2014}. Therefore, the measured proton fluxes may still contain a fraction of anomalous CRs
from inside the heliosphere \citep[][]{Scherer2008}. Second, even if Voyager~1 were in the interstellar space, there is no
guarantee that the measured spectra of CR protons and electrons are representative of the {\em average} Galactic spectra,
because the contribution of local sources (i.e. within parsecs from the Sun) is difficult to quantify. Nevertheless, the
Voyager~1 data provide the only {\em direct} observational constraint presently available on the low-energy spectra of CRs,
and therefore they cannot be discarded. In particular, we use the results of \citet[][]{Stone2013}, obtained from data
collected by Voyager~1 since August 2012, when the spacecraft was at an heliocentric distance of 122~AU.

\begin{figure}\centering
\includegraphics[width=0.85\columnwidth,clip=]{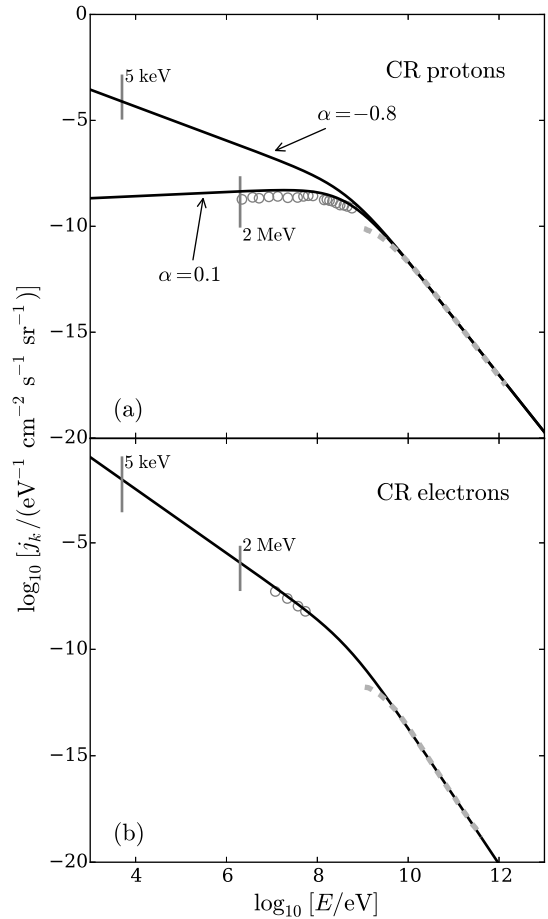}
\caption{Interstellar spectra of CR protons (a) and electrons (b). The black solid lines show the parameterized spectra
given by Eq.~(\ref{IS_spectra}), open circles represent the Voyager~1 data \citep[][]{Stone2013}, gray dashed lines --
the AMS-02 data \citep[][]{Aguilar2014,Aguilar2015}. A combination of the given electron spectrum with the proton spectrum
for $\alpha=-0.8$ and $\alpha=0.1$ is referred to as the CR model $\mathscr{H}$ and $\mathscr{L}$, respectively. The
respective lower energy cutoff $E_{\rm cut}$ ($\simeq5$~keV and $\simeq2$~MeV) is indicated by the gray vertical bars.}
\label{fig1}
\end{figure}

We model the low-energy behavior of the proton and electron spectra with power-law dependencies. To describe a crossover to
the high-energy scalings, we employ the simple analytical expression
\begin{equation}\label{IS_spectra}
  j_{k}(E) = C\frac{E^{\alpha}}{(E+E_{0})^{\beta}}~~\mathrm{eV^{-1} cm^{-2} s^{-1} sr^{-1}},
\end{equation}
where $k=p,e$ and the crossover energy $E_{0}=500$~MeV is the same for both species. Following \citet[][]{Stone2013}, for CR
electrons we adopt a low-energy spectrum slope of $\alpha=-1.5$, while for protons we explore two extreme cases with
$\alpha=-0.8$ and $\alpha=0.1$. The resulting interstellar spectra are presented in Fig.~\ref{fig1}, other parameters for
Eq.~(\ref{IS_spectra}) are listed in Table~\ref{parameters}. Combinations of the two proton spectra with the electron
spectrum are termed below as the model~$\mathscr{H}$ (``High'', $\alpha=-0.8$) and model~$\mathscr{L}$ (``Low'',
$\alpha=0.1$).

\begin{table}[h,t]
\caption{Parameters of interstellar CR spectra, Eq.~(\ref{IS_spectra}).}
\begin{center}
\begin{tabular}{lcccc}
\hline
species ($k$)                        &    $C$   &    $\alpha$    &    $\beta$  & $~\beta-\alpha~$\\
\hline
electrons                            & $2.1\times 10^{18}$  & $-1.5$ & $1.7$ & $3.2$\\
protons (model $\mathscr{H}$)        & $2.4\times 10^{15}$  & $-0.8$ & $1.9$ & $2.7$\\
protons (model $\mathscr{L}$)        & $2.4\times 10^{15}$  & $0.1$  & $2.8$ & $2.7$\\
\hline
\end{tabular}
\end{center}
\label{parameters}
\end{table}

The number densities for the CR species are given by $n_{{\rm CR},k}=4\pi\int_{E_{\rm cut}}^{\infty}dE\: j_{k}(E)/v_{k}(E)$,
where $v_k(E)$ is the velocity for the energy $E$. In order to preserve charge neutrality of CRs, we set $n_{{\rm
CR},p}=n_{{\rm CR},e}$, which yields the lower energy cutoff $E_{\rm cut}\simeq5$~keV and $\simeq2$~MeV for the models
$\mathscr{H}$ and $\mathscr{L}$, respectively. The corresponding energy densities are defined by $\varepsilon_{{\rm
CR},k}=4\pi\int_{E_{\rm cut}}^{\infty} dE\:Ej_{k}(E)/v_{k}(E)$. The total energy density, $\varepsilon_{{\rm
CR},p}+\varepsilon_{{\rm CR},e}$, is dominated by protons and varies between $\simeq0.78$~eV~cm$^{-3}$ (model $\mathscr{L}$)
and $\simeq1.54$~eV~cm$^{-3}$ (model $\mathscr{H}$).

We note that our choice of a single interstellar spectrum for CR electrons and two possible spectra for CR protons is
arbitrary. As discussed above, we adopt the view that the Voyager~1 data represent lower limits to the actual spectra, due
to residual modulation of the interstellar fluxes at the current position of the spacecraft. For simplicity, this remaining
uncertainty is attributed to the protons only: We select the ``minimum'' proton spectrum ($\alpha=0.1$) compatible with the
data, and the ``maximum'' spectrum ($\alpha=-0.8$) providing the upper bound for the available observational data on the CR
ionization rate (see Sec.~\ref{local_spectra}).

\subsection{Local CR spectra}\label{local_spectra}

\begin{figure*}\centering
\includegraphics[width=0.9\textwidth,clip=]{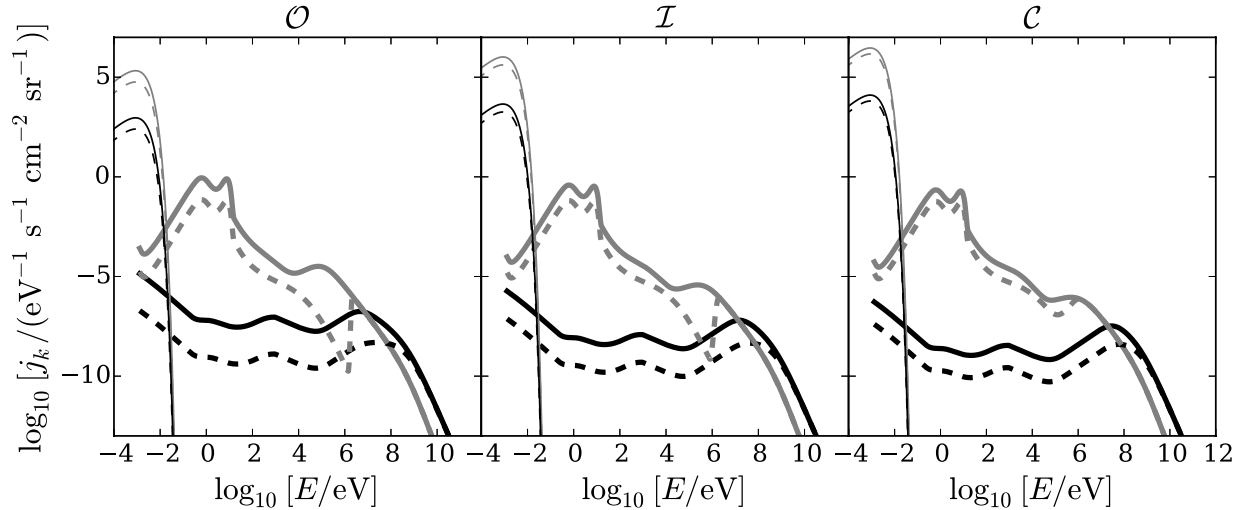}
\caption{Local CR spectra for the model $\mathscr{H}$ (solid lines) and model $\mathscr{L}$ (dashed lines). The three panels
show the results for the three typical characteristic regions of a dense core: ``outer'' (${\cal O}$), ``inner'' (${\cal
I}$), and ``center'' (${\cal C}$). The thin lines represent a cold Maxwellian plasma background, the thick lines are for the
suprathermal propagated CR spectra. The gray and black lines correspond to electrons and protons, respectively. The
presented results are for the case when H$^+$ are the dominant ions in a cold plasma; when heavier ions dominate, the
Maxwellian spectra for ions should be divided by a square root of the corresponding atomic mass number.} \label{fig2}
\end{figure*}

\begin{figure}\centering
\includegraphics[width=0.8\columnwidth,clip=]{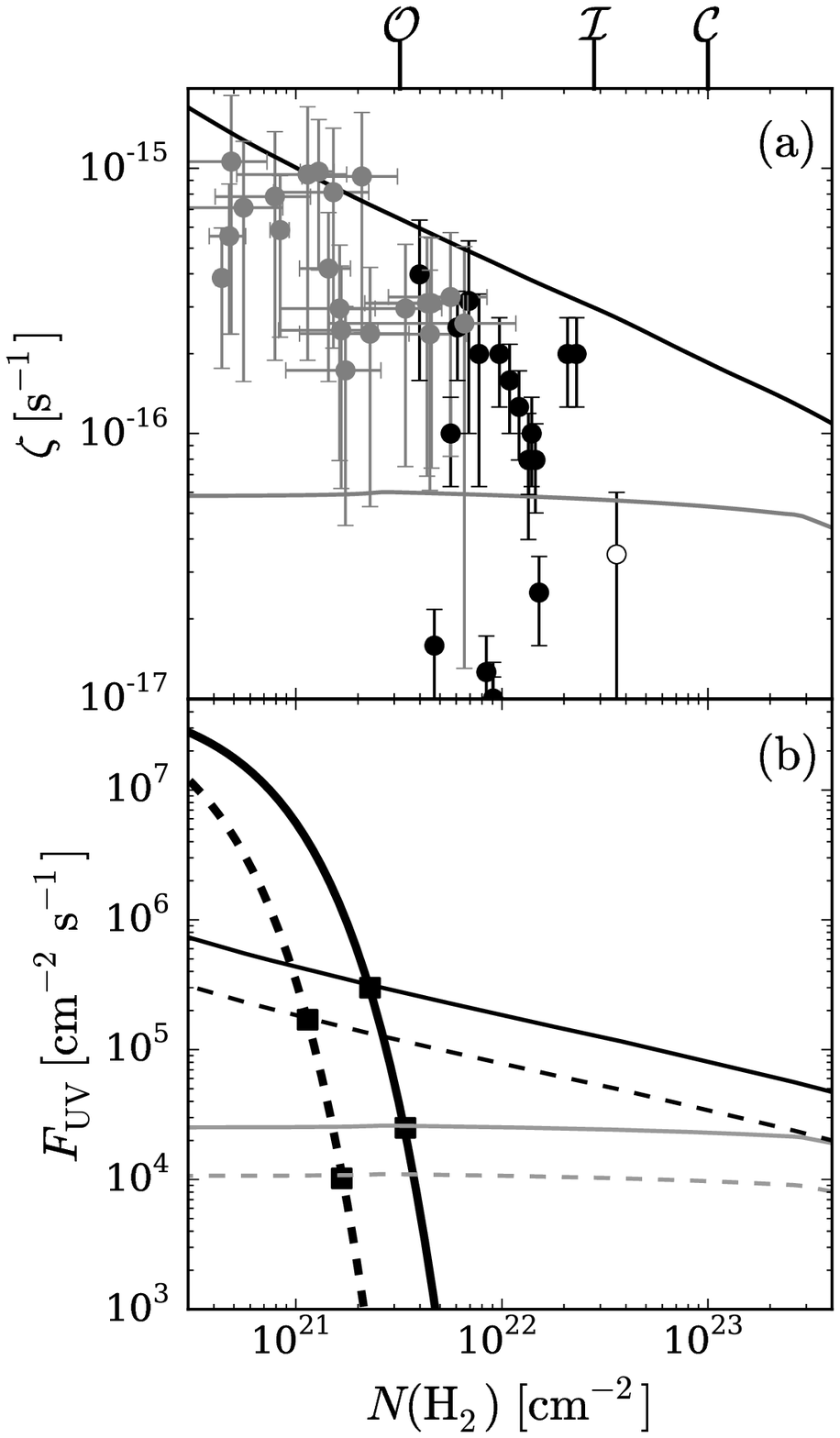}
\caption{(a) CR ionization rate $\zeta$ as a function of the column density of molecular hydrogen $N(\hh)$ for the
model $\mathscr{H}$ (black line) and model $\mathscr{L}$ (gray line), the densities corresponding to the regions ${\cal O}$,
${\cal I}$, and ${\cal C}$ are indicated. The symbols are the observational values
from \citet[][gray solid circles]{Indriolo2012}, \citet[][black solid circles]{Caselli1998}, and
\citet[][empty circle]{Maret2007}. (b) Comparison of the attenuated interstellar flux of UV photons (thick lines) with the
local UV flux generated by CRs (thin lines). The thin black and gray lines represent the CR-generated fluxes for the model
$\mathscr{H}$ and $\mathscr{L}$, respectively. The dashed and solid lines show the results for $R_{\rm V}=3.1$ and 5.5,
respectively.}
\label{fig3}
\end{figure}

In order to concentrate on the charging effects induced by CRs, we adopt an idealized 1D (slab) model of a dense core
embedded in a molecular cloud \citep[for details, see][]{Padovani2009}. We assume that CRs propagate normally to the surface
in straight lines, with a half of the interstellar CR flux incident on each side of the cloud.\footnote{More advanced models
should take into account the fact that molecular clouds are magnetized and CRs gyrate along magnetic field lines, in
addition of being scattered by magnetic fluctuations on the scale of the particle gyroradius. For detailed treatment of
these effects, see \citet[][]{Padovani2011}, \citet[][]{Padovani2013,Padovani2014}, and \citet[][]{Morlino2015}.} This
simple model, neglecting integration over the incidence angles, yields the local spectra that approach the exact results in
the inner core region, while in the outer region the fluxes are only slightly overestimated (by less than 30\%, details will
be presented elsewhere).

For the core we use the density profile of a Bonnor-Ebert sphere \citep[][]{Bonnor1956,Ebert1955}, which typically well
reproduces observations of starless cores \citep[e.g.,][]{Alves2001,Keto2008,Andre2014}. In particular, we consider a
centrally concentrated core, such as L1544, where the volume density within the central 500~AU is
$n(\hh)=2\times10^{7}$~cm$^{-3}$, one of the largest known for starless cores \citep[][]{Keto2010}. The molecular core has a
radius of about 0.1~pc, beyond which the density drops below $n(\hh)=10^{4}$~cm$^{-3}$ and photoprocesses becomes important
\citep[][]{Keto2010}. We further assume that the core is embedded in an envelope of lower density, to simulate the location
of the dense core in a molecular cloud such as Taurus. Such envelope extends up to a distance of $4.4$~pc from the core
center, and the average density within a radius of 4.4~pc is $230$~cm$^{-3}$. With these properties, the cloud has a mass
column density of $144$~M$_{\odot}$~pc$^{-2}$, which is typical of molecular clouds \citep[][]{Roman-Duval2010}.

To demonstrate the extent to which the dust charges in dense molecular clouds are expected to be affected by CRs, we
consider three characteristic regions of the embedded core: the outer boundary [$n(\hh)=10^{4}$~cm$^{-3}$,
$N(\hh)=3.2\times10^{21}$~cm$^{-2}$], the inner core [$n(\hh)=10^{6}$~cm$^{-3}$, $N(\hh)= 2.8\times10^{22}$~cm$^{-2}$], and
also the core center [$n(\hh)=2\times10^{7}$~cm$^{-3}$, $N(\hh)=10^{23}$~cm$^{-2}$]. We shall refer to these regions as
``outer'' (${\cal O}$), ``inner'' (${\cal I}$), and ``center'' (${\cal C}$), respectively.

The CR protons and electrons penetrating the cloud ionize the neutral gas, i.e., produce a local plasma environment whose
properties are determined by the value of $N(\hh)$. On the other hand, the ionization yields a major contribution to the
energy loss of CRs (see Appendix~\ref{app1}), thus modifying the interstellar spectra to the local form
\citep[][]{Padovani2009}. The low-energy CR species as well as the multiple generations of electrons and ions formed in the
ionization avalanche (see Appendix~\ref{app2}) rapidly lose their energy: the characteristic energy loss timescale
$\tau_{\rm stop}$, given by Eq.~(\ref{taustop}), is always much shorter than the timescale of recombination. Therefore, the
steady-state energy distribution of electrons and ions in a dense cloud can be viewed as a superposition of two distinct
parts: a cold Maxwellian peak where electrons and ions accumulate and eventually recombine -- that dominates the total
plasma density, and a suprathermal tail representing the modified (propagated) CR spectrum -- that determines the ionization
rate.

The local steady-state equilibrium is determined by the balance between the CR ionization of H$_2$ and various
recombination processes \citep[see, e.g.,][]{Oppenheimer1974,McKee1989}. A competition between the dissociative
recombination with molecular ions and the radiative recombination with heavy metal ions (in the presence of charge-transfer
reactions), occurring in different core regions, can significantly alter the magnitude of the electron fraction $x_e=
n_e/n(\hh)$, in particular modify the dependence on $n(\hh)$. For the L1544 core, we employ the following interpolation
formula \citep[][their model~3]{Caselli2002}:
\begin{equation}\label{ion_fraction}
    x_e\simeq6.7\times10^{-6}\left(\frac{n(\hh)}{\mathrm{cm^{-3}}}\right)^{-0.56}
    \sqrt{\frac{\zeta}{10^{-17}~\mathrm{s^{-1}}}},
\end{equation}
where $\zeta$ is the CR ionization rate.\footnote{We note that the results presented in Sec.~\ref{distribution} practically
do not depend on the precise form of the formula for $x_e$.} In the following, we assume that the electron density $n_e$ is
equal to the density of all ion species formed in the ionization avalanche (i.e., the contribution of charged grains into
the charge neutrality is negligible, see Sec.~\ref{distribution}).

Figure~\ref{fig2} shows the combination of the propagated spectra and the cold Maxwellian background. The energy $E_{\rm
int}$ at which the intersection between the two curves occurs is practically the same for all considered cases, $E_{\rm int}
\simeq 1.5\times10^{-2}$~eV.

\begin{table}[htdp]
\caption{CR ionization rate $\zeta$ (s$^{-1}$) and plasma electron fraction $x_e$, for three characteristic core regions
(${\cal O}$, ${\cal I}$, and ${\cal C}$) and two CR models ($\mathscr{L}$ and $\mathscr{H}$).}
\begin{center}
\begin{tabular}{lcc}
&$~~~~~~~~~~~\zeta$&\\
\hline
 & model $\mathscr{L}$ & model $\mathscr{H}$\\
\hline
${\cal O}$ & $6.0\times10^{-17}$ & $6.5\times10^{-16}$ \\
${\cal I}$ & $5.6\times10^{-17}$ & $3.0\times10^{-16}$ \\
${\cal C}$ & $5.3\times10^{-17}$ & $1.8\times10^{-16}$ \\
\hline
\end{tabular}
\begin{tabular}{cc}
$~~~~~~~~~~~x_e$&\\
\hline
model $\mathscr{L}$ & model $\mathscr{H}$\\
\hline
$9.4\times10^{-8}$ & $3.1\times10^{-7}$ \\
$6.9\times10^{-9}$ & $1.6\times10^{-8}$ \\
$1.3\times10^{-9}$ & $2.3\times10^{-9}$ \\
\hline
\end{tabular}
\end{center}
\label{ionization}
\end{table}%

Figure~\ref{fig3}a shows the ionization rate as a function of the molecular H$_2$ column density for the two CR models. The
characteristic values of $\zeta$ and the corresponding $x_e$ are summarized in Table~\ref{ionization}. For comparison, the
values of $\zeta$ obtained by integrating the Voyager~1 fluxes down to the lowest measured energy, without any
extrapolation, are $\zeta=1.2\times 10^{-17}$~s$^{-1}$ and $\zeta=2.8\times 10^{-18}$~s$^{-1}$ for protons and electrons,
respectively. These should be considered as lower limits to the interstellar value of $\zeta$.  Notice that the contribution
of CR electrons to the average interstellar CR ionization rate, often neglected in the past, could be significant (although
in general it is smaller than that of protons and heavier nuclei).

We point out that the main positive charge carrier changes across the cloud, because the molecular freeze-out becomes more
efficient toward the core center \citep[see, e.g.,][]{Tafalla_ea2002}. In this paper we do not discuss a complex plasma
chemistry, but instead consider two extreme cases when the dominant ions are either H$^+$ (for the strongly depleted inner
dense regions) or HCO$^+$ (for the outer regions). In Sec.~\ref{mass_dependence} we demonstrate that the plasma composition
has only a minor effect on the obtained results.

\subsection{Local radiation field}\label{local_field}

The local radiation is generated by CRs via the following three (prime) mechanisms: The CR electron bremsstrahlung, the
$\pi^{0}$ decay, and the $\hh$ fluorescence.

We calculate the bremsstrahlung spectrum following \citet[][]{Blumenthal1970}. For all three characteristic core regions,
the resulting photoemission flux from a grain is much smaller than the collection flux of the surrounding cold-plasma ions,
which indicates that the CR bremsstrahlung cannot contribute to dust charging (see Sec.~\ref{distribution:role}). For the
photon spectrum due to $\pi^{0}$ decay, we follow \citet[][]{Kamae2006}. This yields the photoemission flux which is even
smaller than that due to bremsstrahlung and, hence, is negligible, too. Note that for energies larger than the pion
production threshold (280~MeV), both our CR proton spectra coincide and remain unmodified up to
$N(\hh)\sim10^{23}$~cm$^{-2}$.

Finally, following Eq.~(21) in \citet[][]{Cecchi-Pestellini1992}, we compute the $\hh$ fluorescence generated by CRs in the
Lyman and Werner bands. The resulting flux of UV photons $F_{\rm UV}$ (in the energy range between 11.2 and 13.6~eV) can be
approximately calculated assuming that the band excitation rates, being normalized by $\zeta$, do not depend on the shape of
the CR spectrum:\footnote{In fact, the excitation and ionization cross sections have somewhat different dependencies on
energy. For this reason, the normalized excitation rate by CR electrons, obtained with the spectra shown in Fig.~\ref{fig2},
is about 2.2--2.8 times larger than that reported in \citet[][]{Cecchi-Pestellini1992}. However, we were not able to find a
reliable expression for the excitation cross section by protons, and therefore employ the approximate Eq.~(\ref{UV_flux}).
\label{foot2}}
\begin{eqnarray}
    F_{\rm UV}\simeq960\left(\frac{1}{1-\omega}\right)\left(\frac{\zeta}{10^{-17}~\mathrm{s^{-1}}}\right)\hspace{2.5cm}
    \label{UV_flux}\\
\times\left(\frac{N(\hh)/A_{\rm V}}{10^{21}~\mathrm{cm^{-2}mag^{-1}}}\right)\left(\frac{R_{\rm V}}{3.2}\right)^{1.5}
~~\mathrm{cm^{-2}s^{-1}}.\nonumber
\end{eqnarray}
Here, $\omega$ is the dust albedo at ultraviolet wavelengths, and $R_{\rm V}$ is a measure of the slope of the extinction at
visible wavelengths \citep[e.g.,][]{Draine2011Book}. Assuming $\omega=0.5$ and $R_{\rm V}=3.1$
\citep[][]{Cecchi-Pestellini1992},\footnote{According to \citet[][]{Gordon2004}, in dense cores with $R_{\rm V}\gtrsim5$ the
dust albedo may be closer to $0.3$. \label{foot3}} and substituting the typical gas-to-extinction ratio of $N(\hh)/A_{\rm
V}=10^{21}~\mathrm{cm^{-2}\ mag^{-1}}$ we get $F_{\rm UV}\simeq 1830(\zeta/10^{-17}~\mathrm{s^{-1}})$. In
Sec.~\ref{distribution} it is shown that the resulting photoelectric emission from dust grains can significantly exceed the
cold-ion collection. Thus, the strongest radiation field generated by CRs is due to the $\hh$ fluorescence.

The interstellar radiation field is exponentially attenuated in the cloud. The specific intensity at the frequency $\nu$
decreases as $I(\nu)=I_{\rm IS}(\nu)e^{-\tau_{\nu}}$, where $I_{\rm IS}(\nu)$ is the intensity of the interstellar radiation
field according to \citet[][]{Draine2011Book} and $\tau_{\nu}=A_{\nu}/1.086$ is the optical depth. By integrating
$I(\nu)/h\nu$ over $h\nu\geq10~{\rm eV}$ (at lower frequencies, the photoelectric yield from dust grains rapidly falls off,
so the radiation does not contribute to the photoemission), we derive the interstellar photon flux as a function of
$N(\hh)$. The latter is plotted in Fig.~\ref{fig3}b and compared with $F_{\rm UV}$ vs. $N(\hh)$ calculated for the models
$\mathscr{H}$ and $\mathscr{L}$. We see that the photon flux generated by CRs only slightly decreases (model $\mathscr{H}$)
or remains practically constant (model $\mathscr{L}$) in the shown range of $N(\hh)$; even for the maximum value of $R_{\rm
V}=5.5$, the interstellar flux becomes negligible at $N(\hh)\gtrsim3\times10^{21}$~cm$^{-2}$. We conclude that for all three
core regions the radiation field is solely due to CRs.

\section{Charge distribution function of dust}\label{distribution}

The discrete charge distribution $N(Z)\equiv N_Z$ for dust grains of a given size $a$ is normalized to the total
differential dust density at that size, i.e., $\sum_Z N_Z=dn_d(a)/da$. The charge distribution is derived from the detailed
equilibrium of the charging master equation \citep[][]{Draine1987},
\begin{equation}\label{equilibrium}
J_e(Z+1)N_{Z+1}=\left[\sum_iJ_i(Z)+J^{\rm PE}(Z)\right]N_{Z}.
\end{equation}
Here, $J_{e,i}(Z)=J_{e,i}^{\rm M}(Z)+J_{e,i}^{\rm CR}(Z)$ is the electron/ion collection flux, which has the contributions
from cold Maxwellian plasma background (first term) and from suprathermal low-energy part of the CR spectra (second term),
and $J^{\rm PE}(Z)$ is the photoemission flux due to the local radiation field.

We introduce the floating potential $\varphi_Z=Ze^2/a$ of a particle with the charge $Ze$. The collection fluxes of cold
electrons and ions are obtained from the so-called ``orbital motion limited'' (OML) approximation
\citep[][]{Whipple1981,Fortov2005}:
\begin{equation}\label{coll_e}
J_e^{\rm M}(Z)=2\sqrt{2\pi}a^2n_ev_{e}\left\{
\begin{array}{cl}
e^{\varphi_{Z}/k_{\rm B}T}, & Z\leq0;\\
(1+\varphi_{Z}/k_{\rm B}T), & Z\geq0,
\end{array}
\right.
\end{equation}
and
\begin{equation}\label{coll_i}
J_i^{\rm M}(Z)=2\sqrt{2\pi}a^2n_iv_{i}\left\{
\begin{array}{cl}
(1-\varphi_{Z}/k_{\rm B}T), & Z\leq0;\\
e^{-\varphi_{Z}/k_{\rm B}T}, & Z\geq0,
\end{array}
\right.
\end{equation}
where $v_{e,i}=\sqrt{k_{\rm B}T/m_{e,i}}$ are the thermal velocity scales. Note that for the sake of clarity the (minor)
effect of the polarization interaction, omitted in these equations, is discussed later.

The relation between the electron and ion densities is determined from the charge neutrality condition,
\begin{equation*}
    n_e=\sum_in_i+\langle Z\rangle n_d,
\end{equation*}
where $\langle Z\rangle n_d\equiv\int_{a_{\rm min}}^{a_{\rm max}} da\:\sum_ZZN_Z(a)$ is the charge number density carried by
dust. In starless dense molecular clouds studied in this paper, the dust contribution is negligible as long as the ratio
$n(\hh)/\zeta$ is below a certain value -- e.g., for single-size grains of $a=0.1~\mu$m the neutral density should be less
than $\sim10^{11}$~cm$^{-3}$ \citep[][]{Umebayashi1990}, for the MRN distribution \citep[][]{Mathis1977} the approach works
at $n(\hh)\lesssim 10^{7}$~cm$^{-3}$ \citep[][]{Nakano2002}. Therefore, for the calculations below we set $n_e=\sum_in_i$
for all three cloud regions, which implies that our results are independent of details of the grain size distribution and
demonstrate the generic CR effects on dust charging.

We note that in situations where the dust contribution to the charge neutrality is not negligible, e.g., at higher $n(\hh)$,
the distribution of grain charges is determined by the particular form of $dn_d(a)/da$ which is quite uncertain inside dense
molecular clouds \citep[see, e.g.,][]{Weingartner2001b,Kim1994}. Some insights into possible forms of the size distribution
in such environments can be gained from numerical simulations \citep[see, e.g.,][]{Ormel2009}. The role of dust becomes
particularly important in dense cores containing protostellar sources, where small grains (the main carriers of negative
charge) are abundantly produced due to shock shattering \citep[][]{Guillet2011}.

\subsection{``Reference case'': Cold-plasma charging}\label{distribution:ref}

In the usual approach, the dust charging in cold molecular clouds is caused by collection of electrons and ions from a cold
plasma background. The charge distribution in this case is readily obtained by substituting Eqs.~(\ref{coll_e}) and
(\ref{coll_i}) in Eq.~(\ref{equilibrium}):
\begin{eqnarray}
  &\vdots&\nonumber \\
  \frac{N_{+1}}{N_0}&=&\frac1{\sqrt{\tilde m}(1+\tilde\varphi)},\nonumber \\
  \frac{N_{-1}}{N_0}&=&\frac{\sqrt{\tilde m}}{1+\tilde\varphi},\label{reference} \\
  \frac{N_{-2}}{N_{-1}}&=&\frac{\sqrt{\tilde m}e^{-\tilde\varphi}}{1+2\tilde\varphi},\nonumber \\
  &\vdots&\nonumber
\end{eqnarray}
The distribution depends on two dimensionless numbers: The normalized floating potential of the unit charge (or the inverse
normalized temperature) $\tilde\varphi=e^2/ak_{\rm B}T$, and the effective ion-to-electron mass ratio $\tilde m$ (or the
effective atomic mass number $A$) determined by the partial contributions of all ions:
\begin{equation*}
  \tilde\varphi\simeq\frac{1.67}{a_{\mu}T_{10}},\qquad
  \frac{42.8}{\sqrt{\tilde m}}\equiv\frac1{\sqrt{A}}\simeq\sum_i\frac{n_i}{n_e}\frac1{\sqrt{A_i}},
\end{equation*}
where $a_{\mu}$ is in units of $\mu$m, $T_{10}$ is in units of 10~K, and $A_i$ is the atomic mass number of the $i$th ion
species. Both numbers are large: $\tilde\varphi\simeq17$ for $T=10$~K and the largest grains of the MRN distribution
($a\simeq0.1~\mu$m), while $\tilde m=1836$ for a hydrogen plasma. This implies that ({\em i}\/) the abundance of the $Z=-2$
state is exponentially small, and ({\em ii}\/) one can neglect unity in the denominators of Eq.~(\ref{reference}), so the
ratios $N_{\pm1}/N_0\propto aT$ have a simple universal dependence on temperature and dust size. Furthermore, since
$N_{-1}/N_{+1}=\tilde m$, the negatively charged state is at least three orders of magnitude more abundant than the positive
one.

Thus, submicron grains in a cold plasma are either neutral or singly negatively charged. For a given temperature, most of
the dust smaller than $a_{\mu}\simeq1.67/(\sqrt{\tilde m}T_{10})$ is neutral, while larger grains are mostly negatively
charged (for a hydrogen plasma at $T=10$~K, the transition occurs at $a\simeq400$~\AA). This simple scaling holds as long as
$\tilde\varphi\gg1$, i.e., for $a_{\mu}\lesssim T_{10}^{-1}$. Larger grains/aggregates (in the micron-size range, or if the
temperature increases) become multiply negatively charged; the charge distribution remains narrow, with the average charge
$-\langle Z\rangle$ about a few $\tilde\varphi^{-1}$.

\subsection{Effect of CRs}\label{distribution:role}

Figure~\ref{fig2} demonstrates that the CR proton spectrum at lower energies is by many orders of magnitude lower than the
electron spectrum. Therefore, the former gives a negligibly small contribution to the collection flux, as compared to the
flux of CR electrons. Following the approach by \citet[][]{Draine1979}, we calculates the electron collection flux as
\begin{equation}\label{CR flux}
J_e^{\rm CR}\simeq\pi a^2\int_{E_{\rm int}}^{\infty} dE\:4\pi j_e(E)\left[s_e(E)-\delta_e(E)\right],
\end{equation}
where $s_e(E)$ is the sticking coefficient (probability) for impinging CR electrons and $\delta_e(E)$ is the yield of the
secondary electrons (see details in Appendix~\ref{app3}). The lower limit of integration is equal to the intersection energy
$E_{\rm int} \simeq 1.5\times10^{-2}$~eV, at which the CR electron spectra in Fig.~\ref{fig2} cross the Maxwellian curves.

Note that in Eq.~(\ref{CR flux}), under the integral we omitted the OML factor $(1+\varphi_{Z}/E)$ \citep[see,
e.g.,][]{Horanyi1988}, which determines the charge dependence of the flux (and leads to Eq.~(\ref{coll_e}) for the
Maxwellian spectrum). As shown below, the charge distribution is usually concentrated within $|Z|\lesssim3$; since the
magnitude of the floating potential is $e^2/a\sim10$~meV, while the energy of CR electrons contributing to the charging is
$E\sim10$~eV, the OML correction can be safely neglected and hence $J_e^{\rm CR}$ does not depend on $Z$.

\begin{figure*}\centering
\includegraphics[width=0.9\textwidth,clip=]{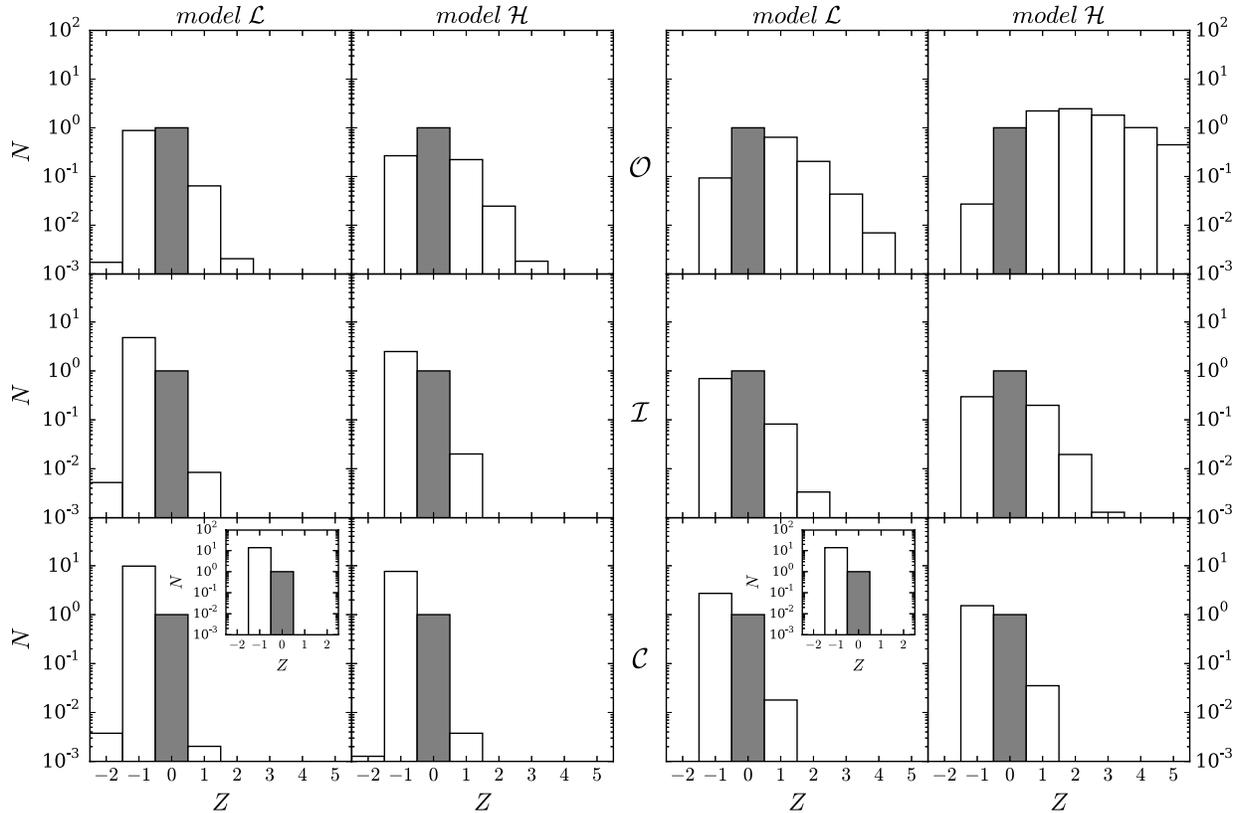}
\caption{Effect of CRs on the dust charge distribution in dense molecular clouds. The distribution function $N_Z$ is
described by Eqs.~(\ref{single1})-(\ref{multiple2}), the results for $\tilde\phi=17$ and $A=29$ (HCO$^+$ ions) are plotted
for different CR models ($\mathscr{L}$ and $\mathscr{H}$) and different characteristic regions (${\cal O}$, ${\cal I}$, and
${\cal C}$) of the core. The left panel shows the results for typical parameters determining the magnitude of $\epsilon_{\rm
PE}$, in the right panel $\epsilon_{\rm PE}$ is increased by a factor of 10. The insets in the lower row depict the
``reference case'', described by Eq.~(\ref{reference}), where the CR effects are neglected. The uncharged state $Z=0$
(shaded) is normalized to unity.} \label{fig4}
\end{figure*}

The photoelectric emission flux from a grain $J^{\rm PE}(Z)$ is determined by the total specific intensity $I(\nu)$ of the
local radiation field, and depends on the photoemission yield $Y(\nu)$ \citep[][]{Weingartner2006} and the absorption
efficiency $Q_{\rm abs}(\nu)$ \citep[][]{Draine2002,Draine2011Book}. As shown in Fig.~\ref{fig3}b, the interstellar UV field
(integrated over the frequency range of $10~{\rm eV}\leq h\nu\leq 13.6~{\rm eV}$) is exponentially attenuated with $N(\hh)$,
and so for all three core regions the local radiation is solely due to the $\hh$ fluorescence generated by CRs. The
resulting photoemission flux can be approximately calculated as
\begin{eqnarray*}
J^{\rm PE}=\pi a^2\int d\nu\:\frac{4\pi I(\nu)}{h\nu}Y(\nu)Q_{\rm abs}(\nu)\hspace{2.5cm}\\
\simeq\pi a^2F_{\rm UV}\langle Y(\nu)Q_{\rm abs}(\nu)\rangle_{\rm UV},
\end{eqnarray*}
where $F_{\rm UV}$ is given by Eq.~(\ref{UV_flux}) and $\langle Y(\nu)Q_{\rm abs}(\nu)\rangle_{\rm UV}$ is averaged over the
Lyman and Werner bands. The latter value may vary for different materials \citep[][]{Weingartner2006}, but this variation is
not very significant; for the calculations below we set $\langle Y(\nu)Q_{\rm abs}(\nu)\rangle_{\rm UV}=0.2$, which
corresponds to a carbonaceous grain of $a\sim0.1~\mu$m. Here, the dependence of $J^{\rm PE}$ on $Z$ is neglected too, since
photoelectrons have a broad energy distribution \citep[roughly limited from above by the energy of the UV photons minus the
work function, see, e.g.,][]{Draine1978} and, hence, the floating potential cannot noticeably affect the photoemission flux.

Thus, the CR effect on the charge distribution is determined by the collection flux of CR electrons $J_e^{\rm CR}$ and the
photoemission flux $J^{\rm PE}$. These fluxes compete, respectively, with the electron and ion fluxes of a cold plasma
background and, hence, their relative magnitude can be conveniently quantified by the dimensionless numbers $\epsilon_{\rm
CR}$ and $\epsilon_{\rm PE}$:
\begin{equation*}
  J_e^{\rm CR}=\epsilon_{\rm CR}J_e^{\rm M}(0),\qquad
  J^{\rm PE} =\epsilon_{\rm PE}\sum_iJ_i^{\rm M}(0).
\end{equation*}
Both numbers do not depend explicitly on $a$ [a weak implicit dependence on dust size is via $s_e(E,a)$ and $\delta_e(E,a)$
for $\epsilon_{\rm CR}$ and via $Y(\nu,a)$ for $\epsilon_{\rm PE}$]. From Eqs.~(\ref{coll_e}) and (\ref{coll_i}) it follows
that $J_e^{\rm M}(0)\propto n_e\sqrt{T}$ and $\sum_iJ_i^{\rm M}(0)\propto n_e\sqrt{T/A}$, and so the relations to the plasma
parameters and the UV flux are given by
\begin{equation}\label{eps_scaling}
    \epsilon_{\rm CR}\propto \frac1{n_e\sqrt{T}},\quad \epsilon_{\rm PE}\propto \frac{F_{\rm UV}}{n_e}\sqrt{\frac{A}{T}}.
\end{equation}

The abundances of the singly charged states are straightforwardly derived from Eq.~(\ref{equilibrium}),
\begin{eqnarray}
\frac{N_{+1}}{N_0}&=&\frac{1+\epsilon_{\rm PE}}{\sqrt{\tilde m}(\tilde\varphi+\epsilon_{\rm CR})},\label{single1}\\
\frac{N_{-1}}{N_0}&=&\frac{\sqrt{\tilde m}(1+\epsilon_{\rm CR})}{\tilde\varphi+\epsilon_{\rm PE}},\label{single2}
\end{eqnarray}
for the multiply charged states the following recurrent relations are obtained:
\begin{eqnarray}
Z\geq2:&\qquad&\frac{N_{+Z}}{N_{+(Z-1)}}=\frac{\epsilon_{\rm PE}}{\sqrt{\tilde m}(Z\tilde\varphi+\epsilon_{\rm CR})},\label{multiple1}\\
-Z\geq2:&\qquad&\frac{N_{-|Z|}}{N_{-(|Z|-1)}}=\frac{\sqrt{\tilde m}\epsilon_{\rm CR}}{|Z|\tilde\varphi+\epsilon_{\rm PE}}.\label{multiple2}
\end{eqnarray}
Following the estimates in Sec.~\ref{distribution:ref}, the unity was neglected with respect to $\tilde\varphi$ in these
equations (as one can see by comparing with Eq.~(\ref{reference})); for the multiply charged states, we also omitted the
exponentially small terms, assuming that $\epsilon_{\rm CR},\epsilon_{\rm PE}\gg e^{-\tilde\varphi}$.

The charge distribution is slightly modified when the polarization interactions are taken into account \citep[see, e.g.,
Eqs.~(3.3) and (3.4) in][]{Draine1987}: In Eqs.~(\ref{single1}) and (\ref{single2}), one has to add the term
$\sqrt{\pi\tilde\varphi/2}$ to the unity in the numerator, and multiply $\tilde\varphi$ in the denominator with 2. For the
multiply charged states (both positive and negative), one needs to replace $|Z|$ in the denominator of
Eqs.~(\ref{multiple1}) and (\ref{multiple2}) with the product $\sqrt{|Z|}(\sqrt{|Z|}+1)$.

One could also take into account the fact that CR electrons (contribution to the collection flux $J_e^{\rm CR}$) have some
finite effective temperature $T_{\rm CR}$. In this case, the collection flux for $Z<0$ and, hence, the r.h.s. of
Eq.~(\ref{multiple2}) has to be multiplied with the OML factor $e^{-|Z|\tilde\varphi_{\rm CR}}$, where $\tilde\varphi_{\rm
CR}=e^2/ak_{\rm B}T_{\rm CR}=\tilde\varphi(T/T_{\rm CR})$. As we pointed out above, the energy of CR electrons contributing
to the charging is of the order of 10~eV, so $\tilde\varphi_{\rm CR}$ is very small ($\lesssim10^{-2}$ for the parameters
used above) and the resulting effect is indeed negligible.

\begin{table}[htdp]
\caption{Dimensionless numbers $\epsilon_{\rm CR}$ and $\epsilon_{\rm PE}$ characterizing the CR effect. The values
correspond to conditions for the left panel of Fig.~\ref{fig4}. For the right panel, $\epsilon_{\rm PE}$ should be
multiplied by 10; for different ions it should be rescaled as $\epsilon_{\rm PE}\propto\sqrt{A}$.}
\begin{center}
\begin{tabular}{lcc}
&$~~~~~~~~~~~\epsilon_{\rm CR}$&\\
\hline
 & model $\mathscr{L}$ & model $\mathscr{H}$\\
\hline
${\cal O}$ & $2.67\times10^{-3}$ & $9.33\times10^{-3}$ \\
${\cal I}$ & $3.35\times10^{-4}$ & $1.34\times10^{-4}$ \\
${\cal C}$ & $8.56\times10^{-5}$ & $4.31\times10^{-5}$ \\
\hline
\end{tabular}
\begin{tabular}{cc}
$~~~~~~~~~~~\epsilon_{\rm PE}$&\\
\hline
model $\mathscr{L}$ & model $\mathscr{H}$\\
\hline
$2.38\times10^{2}$ & $8.71\times10^{2}$ \\
$29.7$             & $74.3$             \\
$7.55$             & $15.0$             \\
\hline
\end{tabular}
\end{center}
\label{CR_numbers}
\end{table}%

Figure~\ref{fig4} illustrates the CR effect on the charge distribution in different regions of the cloud. The left panel
shows $N_Z$ calculated for typical plasma density $n_e$ [from Eq.~(\ref{ion_fraction})] and parameters determining the
magnitude of the CR-generated UV flux $F_{\rm UV}$ [Eq.~(\ref{UV_flux}) with $\omega=0.5$ and $R_{\rm V}=3.1$]. However,
both $n_e$ and $F_{\rm UV}$ are known only approximately: The ionization fraction might be a factor of 2--3 lower than that
given by Eq.~(\ref{ion_fraction}) \citep[see, for example, model~1 of][]{Caselli2002}, while the UV flux is approximately
doubled for $R_{\rm V}\gtrsim 5$ (typical for dark dense clouds, see also Footnote~\ref{foot3}). Furthermore, as pointed out
in Footnote~\ref{foot2}, $F_{\rm UV}$ can exceed the value given by Eq.~(\ref{UV_flux}) due to the difference between the
ionization and excitation cross sections. As one can see from Eq.~(\ref{eps_scaling}), a combination of all these
uncertainly factors may easily increase the value of $\epsilon_{\rm PE}$ by an order of magnitude. Therefore, in the right
panel we show $N_Z$ for $\epsilon_{\rm PE}$ multiplied by 10 (and otherwise the same parameters). The role of CRs in dust
charging becomes particularly evident when the obtained results are compared with the ``reference'' charge distribution
depicted in the insets, where the CR effects are neglected.

From Eqs.~(\ref{multiple1}) and (\ref{multiple2}) we infer that, generally, the asymptotic form of $N_Z$ at large $|Z|$ is a
Poisson distribution $P(Z;\lambda)$: For positive charges, it operates at $Z\gtrsim\epsilon_{\rm CR}/\tilde\varphi$ and is
determined by the Poisson parameter $\lambda_+=\epsilon_{\rm PE}/\sqrt{\tilde m}\tilde\varphi$ (and, analogously, for
negative charges).

Table~\ref{CR_numbers} shows that the collection numbers $\epsilon_{\rm CR}$ are {\it very small} for all considered cases
(at least $10^3$ times smaller than $\tilde\varphi$). Therefore, from Eqs.~(\ref{single1}) and (\ref{multiple1}) we conclude
that positive charges are not affected by CR collection and have a Poisson distribution for all $Z\geq2$; the corresponding
Poisson parameter in the region ${\cal O}$ can be as large as $\lambda_+\simeq0.4$ and $\simeq3.1$ for the models
$\mathscr{L}$ and $\mathscr{H}$, respectively (right panel of Fig.~\ref{fig4}; for the regions ${\cal I}$ and ${\cal C}$,
$\lambda_+\ll1$). On the contrary, the photoemission numbers $\epsilon_{\rm PE}$ are usually {\it large} ($\epsilon_{\rm
PE}\gg\tilde\varphi$, except for the densest ${\cal C}$ case) and the distribution of negative charges is completely
dominated by the state $Z=-1$, whose abundance is given by Eq.~(\ref{single2}); for $|Z|\geq2$ it abruptly decreases as
$N_Z\propto(\sqrt{\tilde m}\epsilon_{\rm CR}/\epsilon_{\rm PE})^{|Z|}$, with $\sqrt{\tilde m}\epsilon_{\rm CR}/\epsilon_{\rm
PE}\lesssim3\times10^{-3}$.

We conclude that in all regions of the cloud, the equilibrium charge distribution is governed by a competition between the
cold-plasma collection (providing negative charging) and the photoemission due to CR-generated UV field (which leads to
positive charging). The effect of photoemission is dominant in the region ${\cal O}$, quite significant in the region ${\cal
I}$ (particularly, in the right panel of Fig.~\ref{fig4}), and still noticeable in the densest region ${\cal C}$. The direct
effect of CRs on dust charging -- the collection of suprathermal CR electrons -- is negligible for all considered
situations.

\subsection{Dependence on the ion mass and grain size}
\label{mass_dependence}

The results, shown in Fig.~\ref{fig4} for HCO$^+$ ions, remain practically unchanged also for other ions: From
Eq.~(\ref{eps_scaling}) it follows that $\epsilon_{\rm PE}\propto\sqrt{\tilde m}$, and therefore the charge distribution for
$Z>0$ [Eqs.~(\ref{single1}) and (\ref{multiple1})] does not depend on $\tilde m$ as long as $\epsilon_{\rm PE}\gg1$, i.e.,
for all considered cases. For $Z<0$ [Eqs.~(\ref{single2}) and (\ref{multiple2})], the distribution is unchanged when
$\epsilon_{\rm PE}\gg\tilde\varphi$. From Table~\ref{CR_numbers} we see that this condition is only violated in the densest
region ${\cal C}$ with the ``typical'' parameters (left panel of Fig.~\ref{fig4}), where $N_{-1}/N_0\propto\sqrt{\tilde m}$.

Using the same consideration, one can also obtain the dependence of $N_Z$ on the grain size: From Eqs.~(\ref{single1}) and
(\ref{multiple1}) we see that, as long as $\tilde\varphi\gg1$, the relative abundance of the positively charged states
varies as $N_Z/N_0\propto \tilde\varphi^{-1} \propto a$. Similar to the dependence on $\tilde m$, the distribution of
negative charges depends on $a$ only in the region ${\cal C}$, where $\epsilon_{\rm PE}\lesssim\tilde\varphi$; in the
regions ${\cal O}$ and ${\cal I}$, the size dependence sets in when the term $e^{-\tilde\varphi}$, neglected in
Eq.~(\ref{multiple2}) [see last Eq.~(\ref{reference})], becomes comparable to $\epsilon_{\rm CR}$. The latter occurs for
sizes $a_{\mu}\geq1.7/ (T_{10}\ln\epsilon_{\rm CR}^{-1})$ which are larger than the upper cutoff $a_{\rm max}$ of the MRN
distribution.

By employing these simple scaling relations and using the examples presented in Fig.~\ref{fig4}, one can easily deduce the
form of $N_Z$ for arbitrary plasma composition and grain size.

\subsection{Implications}

Knowing the charge distribution on dust grains in dense molecular clouds is important for several reasons: The charges
modify the cross sections of the ion accretion on dust, thus critically changing the surface chemistry, influencing the
formation of grain mantles, etc. For the same reason, the charges change the total energy balance of grains and, hence, may
alter the equilibrium temperature. However, the most profound effect can be on the rate of dust coagulation
\citep[][]{Dominik2007}. Let us briefly elaborate on this important point.

Using the derived charge distributions, one can identify the ``optimum'' balance between the positive and negative charges,
for which the rate of dust coagulation is maximized. Conditions for such balance is particularly easy to obtain when the
distribution of positive charges monotonously decreases with $Z$ (i.e., when $\lambda_+<1$). The coagulation rate in this
case is maximized when $N_{+1}\sim N_{-1}$. For $\epsilon_{\rm CR}\ll1$ and $1\ll\tilde\varphi\ll\epsilon_{\rm PE}$ (see
Table~\ref{CR_numbers}), from Eqs.~(\ref{single1}) and (\ref{single2}) we obtain the ``optimum'' relation: $\epsilon_{\rm
PE}^{\rm opt}\sim\sqrt{\tilde m\tilde\varphi}$.

Remarkably, the optimum relation becomes size-independent at late stages of dust coagulation: When the resulting clusters
grow well beyond the upper cutoff size of the MRN distribution, so that $\tilde\varphi\lesssim1$, one has to replace
$\tilde\varphi$ with unity [neglected in Eqs.~(\ref{single1}) and (\ref{single2})]. This yields
\begin{equation*}
    \epsilon_{\rm PE}^{\rm opt}/\sqrt{\tilde m}\sim1.
\end{equation*}
Furthermore, since $\epsilon_{\rm PE}\propto\sqrt{\tilde m}$, the coagulation optimum does not depend on the plasma
composition either and, hence, is universal. As one can see from Table~\ref{CR_numbers}, the optimum is attained somewhere
between the regions ${\cal O}$ and ${\cal I}$, which corresponds to the densities $n(\hh)$ between $\sim10^4$~cm$^{-3}$ and
$\sim10^6$~cm$^{-3}$.

The existence of the coagulation optimum is in striking contrast with the case of pure plasma charging, where growing
negative charges on clusters inhibit their further coagulation \citep[][]{Okuzumi2009}. Thus, CRs can provide ideal
conditions for a rapid dust coagulation in cold dense molecular clouds. We will discuss this important effect in a separate
paper.

\section{Conclusions}

The aim of this paper is to demonstrate that CRs strongly affect charging of dust grains in cold dense molecular clouds.

We calculated the local (propagated) CR spectra for three characteristic regions of a dense core, and investigated the two
mechanisms of dust charging that have been ignored so far: collection of suprathermal CR electrons and protons by grains
(adding to the cold Maxwellian plasma collection), and photoelectric emission from grains due to the CR-generated UV field.
While the former mechanism turns out to be always negligible, the photoemission is shown to dramatically modify the charge
distribution for submicron grains in the almost entire cloud (as compared to the ``reference case'' of cold-plasma
charging). The competition between the cold-plasma collection (producing, primarily, singly charged negative grains) and the
photoemission (resulting in positive charging) significantly broadens the charge distribution.

The relative magnitude of the CR-induced photoemission is quantified by the dimensionless number $\epsilon_{\rm PE}$. This
number depends on several physical parameters, some of them being only approximately known. The mains sources of uncertainty
are the values of the ionization fraction $x_e$ and the CR-induced UV flux $F_{\rm UV}$: As we pointed out in
Sec.~\ref{distribution:role}, the uncertainty in $x_e$ is mainly due to competition between different recombination
processes, while $F_{\rm UV}$, in turn, is determined by (approximately known) dust albedo, extinction slope, and
photoemission yield, as well as by (partially unknown) $\hh$ excitation rates. To take all these factors into account, in
Fig.~\ref{fig4} we presented two sets of plots showing the dust charge distribution: The left and right panels correspond to
the expected ``typical'' and ``maximum'' values of $\epsilon_{\rm PE}$ and, thus, demonstrate the extent to which the
integral effect of uncertainties may affect the final results.

Our results have several important implications. The shown modification of the grain charge distribution considerably
changes the rates of ion accretion on dust, which in turn can critically change the surface chemistry, alter the total
energy balance of grains, influence the formation of icy mantles, etc. These CR effects are particularly strong in the outer
regions of the core, where the charge distribution is dominated by positive grains and, hence, the accretion of negatively
charged ions should be drastically reduced.

The most profound effect of CRs is expected to occur for the rate of dust coagulation: When the cold-plasma collection is
the only charging mechanism operating in a cloud, the average (negative) dust charge increases proportionally to the size.
Therefore, the growing Coulomb repulsion inhibits coagulation of larger ($\gtrsim1~\mu$m) aggregates. Here we showed that
the competition between the cold-plasma collection and photoemission can create approximately equal abundance of positively
and negatively charged dust, providing ``optimum'' conditions for coagulations. The derived optimum is size-independent for
large dust, which enables the growth of big aggregates.

The presented results are obtained assuming that charged dust does not affect the overall charge neutrality in a cloud.
Although this assumption can be violated for sufficiently dense regions, our approach can be straightforwardly extended to
this case as well, provided the dust size distribution is known. Furthermore, the described effects of CRs are of generic
nature, and are expected to operate not only in dense molecular clouds but also in the upper layers and the outer parts of
protoplanetary discs, where mutual sticking of dust aggregates is the essential process toward planetesimal formation.

\vspace{0.5cm} The authors would like to thank the referee A. Jones for providing helpful and constructive comments and
suggestions. The authors also acknowledge Malcolm Walmsley for critical reading of the manuscript and helpful comments. MP
acknowledges the support of the OCEVU Labex (ANR-11-LABX-0060) and the A*MIDEX project (ANR-11-IDEX-0001-02) funded by the
``Investissements d'Avenir'' French government programme managed by the ANR. MP and DG also acknowledge the support of the
CNRS-INAF PICS project ``Pulsar wind nebulae, supernova remnants and the origin of cosmic rays''. PC acknowledges support
from the European Research Council (ERC, project PALs 320620).

\appendix

\section{Appendix A\\ Coulomb loss functions}\label{app1}

\begin{figure}\centering
\includegraphics[width=0.4\columnwidth,clip=]{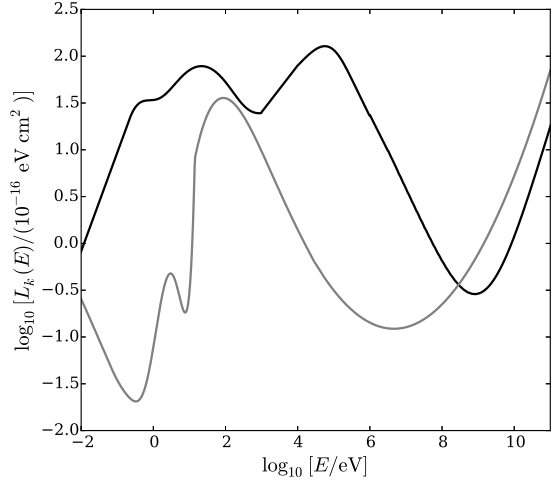}
\caption{Energy loss functions for protons (black) and electrons (gray). The ionization fraction $x_{e}=10^{-7}$
has been assumed to include Coulomb losses.}
\label{fig1app}
\end{figure}

In order to combine the propagated CR spectra obtained in \citet[][]{Padovani2009} with the cold Maxwellian plasma
background, the energy loss functions $L_k(E)$ \citep[plotted in Fig.~7 of][]{Padovani2009} have to be extended to lower
energies (down to about 1 meV), to include Coulomb losses that dominate in this energy range. The Coulomb loss term,
$L_k^{\rm C}(E)$, is parameterized for protons by \citet{Schlickeiser2002} as
\begin{equation*}
    L_p^{\rm C}(E)\simeq4.8\times10^{-10}x_{e}E^{-1}~~\mathrm{eV~cm^{2}},
\end{equation*}
where $x_{e}$ is the ionization fraction, the proton energy $E$ (in eV) is supposed to exceed $\sim 3k_{\rm B}T$. For
Coulomb electron losses, we used the analytic fit by \citet[][]{Swartz1971},
\begin{equation*}
    L_e^{\rm C}(E)\simeq 5.7x_{e}E^{-0.94}~~\mathrm{eV~cm^{2}}.
\end{equation*}
Dense cores have typical temperatures of about 10~K, so $k_{\rm B}T\simeq9\times10^{-4}$~eV. Figure~\ref{fig1app} shows the
energy loss functions from \citet[][]{Padovani2009} with the Coulomb terms included, assuming an average ionization fraction
of $x_e=10^{-7}$.

\section{Appendix B\\ Specific intensity of secondary electrons}\label{app2}

After crossing a column density $dN=n\, dx$ in medium of density $n$, an electron of energy $E$ and velocity $v_e$ has lost
an energy $dE=-L_e(E)\, dN$, where $L_e(E)$ is the energy loss function of electrons,
\begin{equation}\label{loss}
    L_e(E)=-\frac{1}{n}\frac{dE}{dx}=-\frac{1}{nv_e}\frac{dE}{dt},
\end{equation}
(and similarly for protons). The column density $N_{\rm stop}$ required to stop an electron of the initial energy $E$,
\begin{equation*}
    N_{\rm stop}(E)=\int_0^{E}\frac{dE'}{L_e(E')},
\end{equation*}
has a constant value $N_{\rm stop}(E)\sim10^{18}$~cm$^{-2}$ up to $E\sim1$~keV \citep[see Fig.~8 in][]{Padovani2009}, and
then increases weakly with energy. This short range of secondary electrons, in comparison with typical column densities of
molecular clouds ($\sim10^{22}-10^{23}$~cm$^{-2}$) justifies a local treatment of ionization. In this ``on-the-spot''
approximation, the stopping time of secondary electrons is
\begin{equation}\label{taustop}
    \tau_{\rm stop}(E)\simeq \frac{E}{n v_e L_e(E)},
\end{equation}
and their stopping range is $\sim v_e \tau_{\rm stop}\simeq E/[nL_e(E)]$.

Assuming isotropy, the number of first-generation secondary electrons of energy $E$ (per unit energy, volume and time) is
calculated as
\begin{equation*}
    \frac{d{\cal N}_e^{\rm sec}}{dE dV dt}=4\pi n \int_{I+E}^\infty dE'\: j_k(E') \frac{d\sigma^{\rm ion}_k}{dE}(E,E'),
\end{equation*}
where $I=15.6$~eV is the ionization potential of H$_2$, $j_k(E')$ is the intensity of primary species $k$, and $d\sigma^{\rm
ion}_k/dE$ is the corresponding differential ionization cross section. The number of secondary electrons produced per unit
energy and volume is then
\begin{equation*}
    \frac{d{\cal N}_e^{\rm sec}}{dE dV}\simeq \frac{d{\cal N}_e^{\rm sec}}{dE dV dt}\tau^{\rm stop}=\frac{4\pi E}{v_e
L_e(E)} \int_{I+E}^\infty dE'\: j_k(E') \frac{d\sigma^{\rm ion}_k}{dE}(E,E').
\end{equation*}
This quantity is related to the specific intensity of secondary electrons $j_e^{\rm sec}(E)$ (number of electrons per unit
energy, area, time and solid angle) by $d{\cal N}_e^{\rm sec}/dE\, dV=(4\pi/v_e)j_e^{\rm sec}$,
which finally yields
\begin{equation}\label{sec}
    j_e^{\rm sec}(E)\simeq\frac{E}{L_e(E)} \int_{I+E}^\infty dE'\: j_k(E') \frac{d\sigma^{\rm ion}_k}{dE}(E,E').
\end{equation}
Equation~(\ref{sec}) is iterated, to compute intensities for the next generations of secondary electrons.

\section{Appendix C\\ Electron sticking probability and secondary emission yield}\label{app3}

Following \citet[][]{Draine1979}, we set the sticking probability $s_{e}(E)$ equal to unity if the electron stopping range
$R_{e}(E)$ (in the dust material) is smaller than $4a/3$, otherwise $s_{e}(E)=0$. The transition energy varies with dust
size as $\propto a^{2/3}$; for sub-micron grains, $a\sim0.1~\mu$m, the transition occurs at $E\sim3$~keV. The electron
trapping by small (PAH) grains was calculated by \citet[][]{Micelotta2010}.

The secondary emission yield $\delta_{e}(E)$ is obtained by averaging over the velocity distributions of the emitted
electrons, which are believed to be broad (non-Maxwellian), decaying as $E^{-1}$ at large energies \citep[][]{Draine1979}.
For a positively charged dust the explicit dependence on $\varphi_{Z}$ is approximated by $\delta_{e}\propto1/\sqrt{1
+(\varphi_{Z}/E_*)^2}$, where $E_*\sim3$~eV. This factor can be neglected using the same reasoning as for the OML correction
factor omitted in Eq.~(\ref{CR flux}). For the dependence on $E$ we employ the Sternglass formula \citep[][]{Horanyi1988},
\begin{equation*}
    \frac{\delta_e(E)}{\delta_e^{\rm max}}=\frac{E}{E_{\rm max}}\exp\left(2-2\sqrt{\frac{E}{E_{\rm max}}}\right),
\end{equation*}
where the value of $\delta_e^{\rm max}$ is typically between 1.5 and 2.5, and $E_{\rm max}=0.2-0.4$~keV
\citep[][]{Draine1979}. Note that $\delta_{e}(E)$ has a finite-size correction at large $E$, which increases the yield by a
factor of $\simeq1.6$.

\bibliographystyle{apj}
\bibliography{refs}

\end{document}